\pdfoutput = 1

\documentclass[doublecol]{epl2}

\title{On the Use of Topological Features and Hierarchical Characterization for Disambiguating Names in Collaborative Networks}
\shorttitle{Complex Networks for Name Disambiguation} 

\author{Diego R. Amancio\inst{1} \and Osvaldo N. Oliveira Jr.\inst{1} \and Luciano da F. Costa\inst{1}}
\shortauthor{D. R. Amancio \etal}

\institute{
  \inst{1} Institute of Physics of S\~ao Carlos \\
	University of S\~ao Paulo, P. O. Box 369, Postal Code 13560-970 \\
	S\~ao Carlos, S\~ao Paulo, Brazil \\ \ \\ \\
}


\usepackage{color}
\definecolor{greencolor}{rgb}{0,0.5,0.2}
\definecolor{redcolor}{rgb}{1.0,0.,0.}
\definecolor{bluecolor}{rgb}{0,0.,1.}
\definecolor{greycolor}{rgb}{.5,.5,.5}

\usepackage[nearskip,caption=false,font=footnotesize]{subfig}
\usepackage[pdftex]{hyperref}

\abstract
{
Many features of complex systems can now be unveiled by applying statistical physics methods to treat them as social networks. The power of the analysis may be limited, however, by the presence of ambiguity in names, e.g. caused by homonymy in collaborative networks. In this paper we show that the ability to distinguish between homonymous authors is enhanced when longer-distance connections are considered, rather than looking at only the immediate neighbors of a node in the collaborative network. Optimized results were obtained upon using the 3rd hierarchy in connections. Furthermore, reasonable distinction among authors could also be achieved upon using pattern recognition strategies for the data generated from the topology of the collaborative network. These results were obtained with a network from papers in the arXiv repository, into which homonymy was deliberately introduced to test the methods with a controlled, reliable dataset. In all cases, several methods of supervised and unsupervised machine learning were used, leading to the same overall results. The suitability of using deeper hierarchies and network topology was confirmed with a real database of movie actors, with the additional finding that the distinguishing ability can be further enhanced by combining topology features and long-range connections in the collaborative network.
}

\begin{document}

\maketitle

\section{Introduction}

The e-Science paradigm
may be exploited to transform the tremendous amounts of data electronically available into useful knowledge in varied fields. In science and technology, for example, large databases include citation networks~\cite{cit5}, journals databases~\cite{pre2}, arXiv\footnote{http://www.arXiv.org}, CiteSeer\footnote{http://citeseerx.ist.psu.edu}, DBLP\footnote{http://www.informatik.uni-trier.de/~ley/db/}, Web of Science\footnote{http://apps.isiknowledge.com} and Google Scholar\footnote{http://scholar.google.com}, whose analysis may assist in the decision-making process of funding agencies and academic institutions. Citation networks, in particular, have been studied with a variety of purposes, e.g. identifying the most relevant papers in a survey and quantifying the impact of journals, conferences, researchers and institutions. The applicability of these databases may be hampered, nevertheless, if they are not accurate or if they contain ambiguities. For scientific databases, two major problems appear in lists of authors of scientific articles: i) the same author may be referenced in different ways and ii) distinct authors may have identical names, which is especially important for Chinese and Korean researchers~\cite{chao}.

Several methods have been used to resolve ambiguities of authors' names in scientific papers, which is a task akin to several other problems, such as matching~\cite{sim2} and duplicate detection~\cite{sim4}. These methods are mostly based on text mining
~and on natural language processing~\cite{manning}, because researchers are believed to be fairly characterized by their research field, so that textual similarity measures are able to cluster together manuscripts authored by the same scientist. {The list of co-authors has also been used as a criterion for disambiguation~\cite{ref1,ref2} since authors tend to keep a specific collaboration group.} Of lesser importance are the criteria based on the journal name~\cite{prob}, language of the manuscript~\cite{prob}, authors' affiliation~\cite{prob}, self-citations~\cite{activia} and source URL metadata~\cite{activia}.

Another approach to detect and repair inconsistencies in databases is to represent them as complex networks~\cite{cit5,net8}. In this paper, we use concepts and metrics of networks to distinguish between authors represented by the same alias in a collaborative network. The network was retrieved from arXiv\footnote{http://www.arXiv.org}, where homonymy was deliberately introduced to have a reliable dataset, and distinction was made with two approaches. In the first, we employed deeper local hierarchies~\cite{hier1} for analyzing the connectivity of the collaborative network, while in the second topological features of the network were used. The data generated from the analysis were treated with projection techniques~\cite{lsa,pca} to reduce dimensionality and pattern recognition methods were used in distinguishing authors. The two methodologies, with deeper hierarchies and topological features, were combined to disambiguate actors' names in the IMDb\footnote{http://www.imdb.com} database.

\section{Methodology}

\subsection{Databases}

Two databases were used, the first of which is a set of preprint manuscripts from the arXiv repository (see footnote 1). The articles were retrieved using the keywords ``complex network'' or "scale free". The second database was retrieved from the IMDb repository (see footnote 6). Only movies released after the year $2000$ were considered. {Details concerning both databases are given in the Supplementary Information\footnote{The Supplementary Information is available from \url{https://dl.dropbox.com/u/2740286/eplSI9mai.pdf}.} (SI).}

\subsection{Network Formation}

\begin{figure*}[!Htb]
\centering
\subfloat[]
{\includegraphics[scale=0.65]{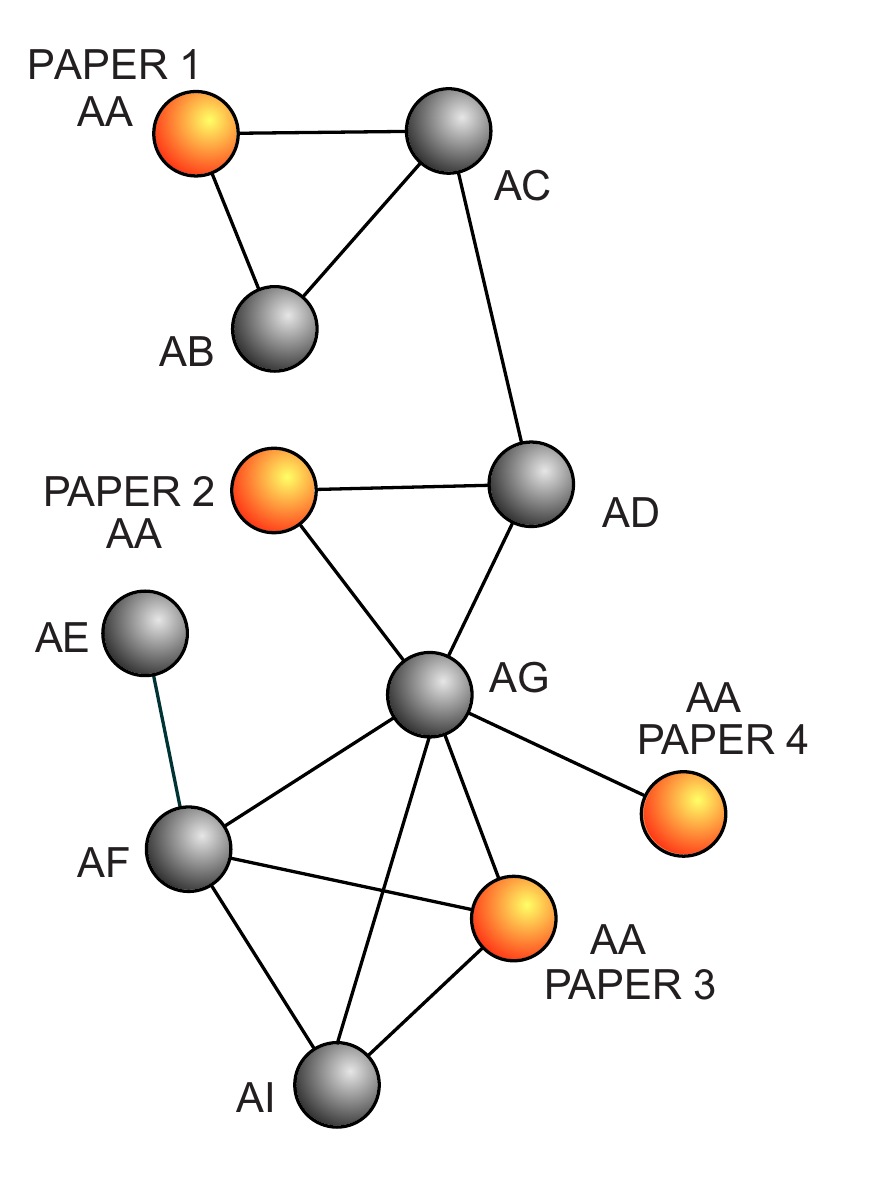}\label{netExample_a}}%
\subfloat[]
{\includegraphics[scale=0.5]{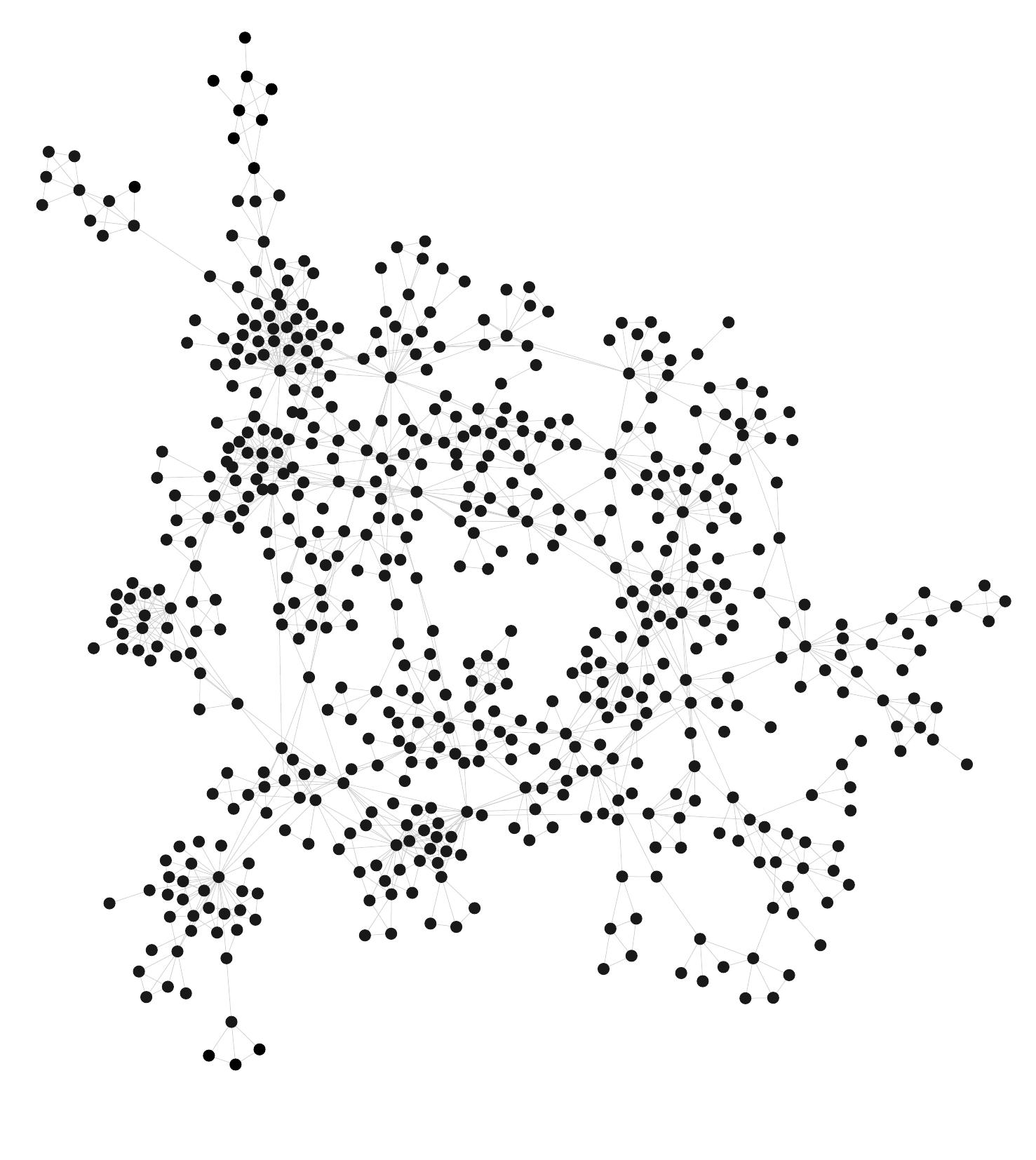}\label{netExemple_b}}%
\caption{(a) Example of a collaborative network built from a fictitious list of authors and (b) giant component of the collaborative network of a fraction of the arXiv repository. AX stands for Author X. In (a), the following authorship of papers was considered: paper 1 (AA, AB and AC), paper 2 (AA, AG and AD), paper 3 (AA, AF, AG and AI), paper 4 (AA and AG), paper 5 (AC and AD), paper 6 (AE and AF) and paper 7 (AF and AE).
}
\end{figure*}

Collaborative networks were generated using the two databases, in which the nodes represented the authors or actors, being linked if they co-participated in a paper or movie. The process of building the collaborative network of authors is illustrated in Figure~\ref{netExample_a} for a small network with $7$ fictitious papers (see caption), while Figure \ref{netExemple_b} shows the giant component of the arXiv collaborative network for the subject of ``complex networks''. In the fictitious papers shown in the figure, the aim is to disambiguate authors with the same name \emph{AA}. {Note that it is necessary to represent the author of interest as different nodes in the network because the disambiguation is performed at the paper level. Anyway, one does not need to create a completely different network every time one wants to disambiguate a specific author. A common network reflecting all collaborations in the database could be maintained and only a few nodes and edges would be added/removed from the common network for the analysis of each new particular author. Thus the disambiguation process would fundamentally depend on the number of papers co-authored by the author under analysis.}

The strength of the connection between vertices $i$ and $j$ using the weight $w_{ij}$ is:
\begin{equation}
w_{ij} = \sum_{\rho \in \Pi} \frac{ \delta_{ij\rho} }{ \|\rho\| }, \; \; \textrm{where}
\label{eq:formulaDiego}
\end{equation}
\begin{equation} \label{cond}
\delta_{ij\rho} = \left\{
\begin{array}{rl}
1 & \textrm{if  $i$ and $j$ appear in paper $\rho$,} \\
0 & \textrm{otherwise }\\
\end{array} \right.
\end{equation}
$\Pi$ represents the set of all papers in the database and $\|\rho\|$ is the number of authors of a given paper $\rho$. The weight was divided by $\|\rho\|$ to take into account the finding that relationships among few authors are usually stronger than those involving several authors~\cite{unsupd}. The weight of the links is not shown in Figure~\ref{netExample_a}, but its computation is straightforward. For instance, the weight for the link between \emph{AB} and \emph{AC} is $1/3$ while that for \emph{AE} and \emph{AF} is $1$ ($1/2$ from paper $6$ plus $1/2$ from paper $7$).

    \subsection{Characterization of Entities Through Connectivity Analysis}
    \label{cnmeasurements}

    In the strategy based on co-authorship, each occurrence of an ambiguous entity is characterized by relations of co-participation in the same paper/movie. Let $e$ be an ambiguous entity and let $\overrightarrow{v_e}$ be the vector describing the co-authorship features of $e$. Each element $i$ in $\overrightarrow{v_e}$ represents one of the possible entities in the database. As such, if $i$ and $e$ appear in the same document, then $\overrightarrow{v_e}(i) = 1$. Otherwise, $\overrightarrow{v_e}(i) = 0$. In order to reduce the complexity of the problem, two techniques to reduce dimensionality were used: principal component analysis~\cite{pca} (PCA) and latent semantic analysis~\cite{lsa} (LSA). Because it performed better than PCA in the experiments, all of the results reported here were obtained with LSA. {Both techniques are described in the SI.}

    \subsection{Characterization of Entities Through Topological Analysis}
    \label{cnmeasurements}

    In addition to the strategy based on co-authors, we evaluated the suitability of the local topological structure for disambiguating names. The measurements used were: degree $k$, which quantifies the number of links; strength $s$, which quantifies the sum of the weights of links; clustering coefficient $C$, which measures the density of links around the node of interest; average degree $\langle k_n \rangle$ and strength $\langle s_n \rangle$ of immediate neighbors; and the standard deviations $\sigma_{k_n}$ and $\sigma_{s_n}$ of degree and strength of neighbors, respectively. Further details on complex networks measurements are given in Refs.~\cite{net8,beyond}.

    \subsection{Hierarchical Characterization}

    Inspired by studies showing that the expansion of local analysis for further neighbors allows better characterization of networks~\cite{amancio}, we introduced the hierarchical analysis in the characterization of collaborative networks. When the hierarchy of a given node is expanded, all of its neighbors $v_n$ are lumped into a single new node $v_h$. As a result, if any other node $v_i$ of the network was connected to $v_n$ before the expansion, then afterward $v_h$ will be connected to $v_i$. In our experiments, the networks were expanded twice, therefore three hierarchies were generated. Details on the hierarchical characterization in complex networks are given in Refs.~\cite{hier1,prlluk}.

    \subsection{Pattern Recognition Techniques}

    Pattern recognition techniques that induce classifiers from the training set were used in the disambiguation task, employing features extracted from the analysis of connectivity and topology of the collaborative networks. The quality of the results was then evaluated using the 10-fold cross-validation technique~\cite{bishop}, which was chosen because it is robust since the training set is always different from the evaluation set. Thus, it prevents that overfitted inductors take high values of accuracy rate. We used methods belonging to the paradigms, namely, supervised and unsupervised techniques. In the former, a function is inferred upon the labeled training data. The four techniques used were: C4.5 algorithm~\cite{bishop}, which generates trees based on the gain provided by each feature; Naive Bayes algorithm~\cite{bishop}, which uses the Bayes theorem; $k$ nearest neighbor algorithm~\cite{bishop} ($k$NN), which classifies an external unknown instance according to the most similar instance of the training database in a normalized space including all features; and {\it RIPPER} algorithm, which generates a set of explicit rules to classify new instances. In the unsupervised methods, one does not know in advance which element belongs to each class, what is known is that a given pair of names belongs to the same entity. The techniques used were: $k$-means~\cite{bishop}, Expectation Maximization (EM)~\cite{bishop}, single linkage~\cite{jaindubes}, complete linkage~\cite{jaindubes}, average linkage~\cite{jaindubes} and Ward's linkage~\cite{jaindubes}. After the classification phase, two quality indicators were employed to assess the performance: the rate of instances correctly classified and the f-measure~\cite{fscore}, which represents a balance between precision and recall of correctly classified instances. The algorithms, the cross-validation technique and the f-measure are described in the SI.

    The reasons why several supervised and unsupervised machine learning methods were used are related to ensuring robustness of the data analysis, especially because we shall show that the overall conclusions are independent of the pattern recognition method.

    \section{Results and Discussion}

    \subsection{Disambiguation based on the connectivity of authors}

    In this strategy we used a set of $N = 1,842$ features, where $N$ is the number of authors of the arXiv database. Because the data including the $N$ features for the various homonymous authors had a high dimension, we employed PCA and LSA to reduce the dimension. Then we used the pattern recognition strategies mentioned in the methodology. The analysis of the immediate neighborhood of authors in the collaborative network allows for distinction of homonymous authors, with the overall accuracy increasing when deeper hierarchies were used. Figure \ref{fig.2} shows the f-measure obtained with the 3rd hierarchy for the arXiv network, which indicates that the performance decreased with the number of homonymous authors, as expected, and this applies to all algorithms tested (see also Figure S1 of the SI). The superior performance of the analysis considering the 2nd and 3rd hierarchies is depicted in Figure \ref{fig.6}, which shows the percentage in which each hierarchy achieved the best performance (a statistical analysis of the figure is provided in Table S1 of the SI). These results are consistent with the finding in a previous study where the use of higher hierarchies improved the local characterization of networks~\cite{amancio}. Hierarchies higher than 3 were not attempted owing to the high computational cost. Nevertheless, the performance is unlikely to increase considerably for deeper hierarchies, and should indeed be expected to decrease if very high hierarchies were used because more information might be lost than gained~\cite{amancio}.

    \begin{figure}[h]
  \centering
    \subfloat[]
    {{\includegraphics[angle=0, width=0.24\textwidth]{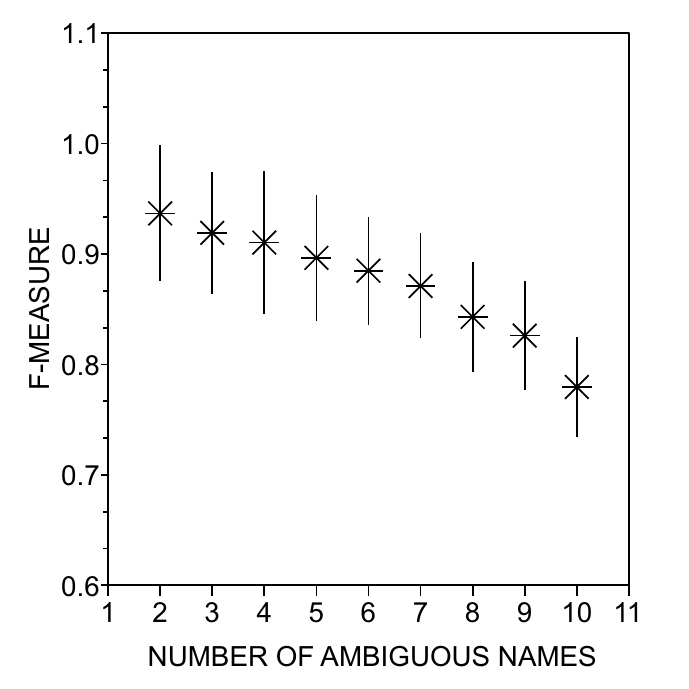}\label{fig:Regular-Lattice-lambda0}}}%
    \subfloat[]
    {{\includegraphics[angle=0, width=0.24\textwidth]{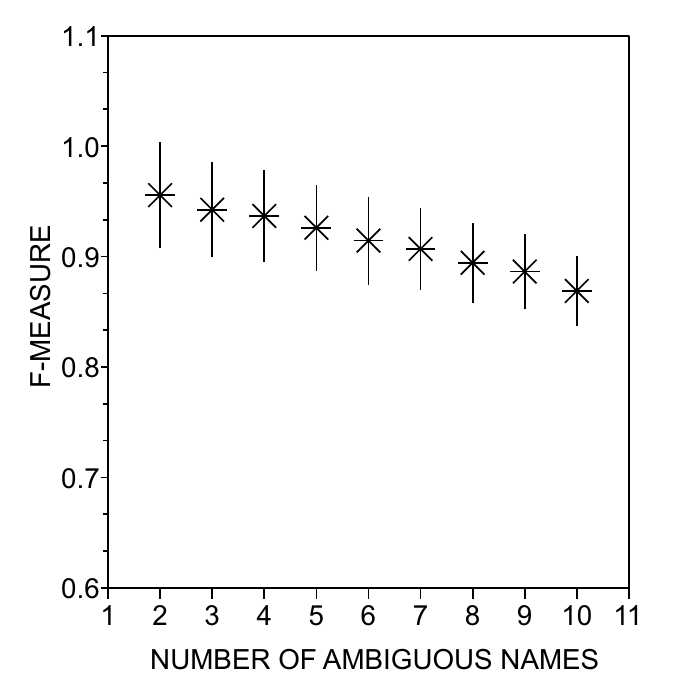}\label{fig:Regular-Lattice-lambda05}}}\\
    \caption{\label{fig.2} f-measure for the disambiguation task from the analysis of connectivity of the collaborative network using the 3rd hierarchy. The algorithms used were (a) C4.5 and (b) $k$NN-1. In all cases, the ability to distinguish among authors decreased as the number of ambiguous entities increased.}
    \label{fig.2}
    \end{figure}



\begin{figure}[h]
    \begin{center}
        \includegraphics[width=0.49\textwidth]{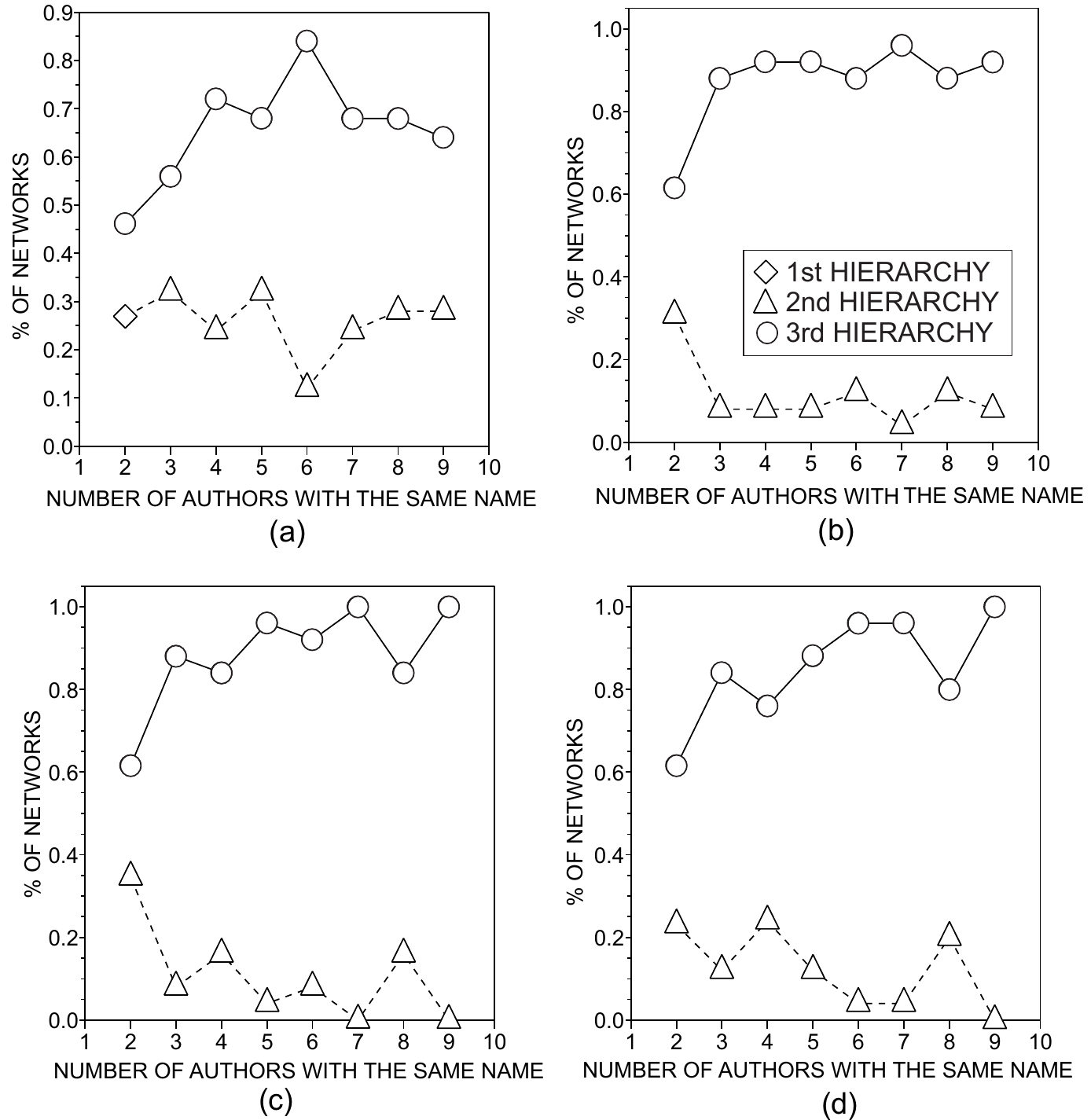}
    \end{center}
    \caption{\label{topologia} Percentage of cases where each hierarchy achieves the higher value of the f-measure for the following algorithms: (a) complete; (b) Ward; (c) K-Means; (d) Expectation Maximization. Note that, in most cases, the 3rd hierarchy outperforms the 2nd and 1st hierarchies. A similar behavior was observed for the other algorithms.}
    \label{fig.6}
\end{figure}

{The disambiguation process in collaborative networks may depend on the edge density of the collaborative networks. If everybody is connected to everybody else, then the disambiguation process tends to deteriorate and other factors in addition to co-authorship relations should be included to discriminate authors' names. Fortunately, in practice, collaborative networks are organized in communities so that the clustering is high only within communities. As such, the high clustering within communities is desirable when ambiguous authors belong to distinct communities. The higher the clustering the smaller the number of external links will be. As a result, authors' co-authorship patterns will be quite distinct provided they belong to different communities.}

\subsection{Topological features used in distinguishing authors}

\begin{figure} [!Htb]
    \centering
    \subfloat[]
    {{\includegraphics[angle=0, width=0.24\textwidth]{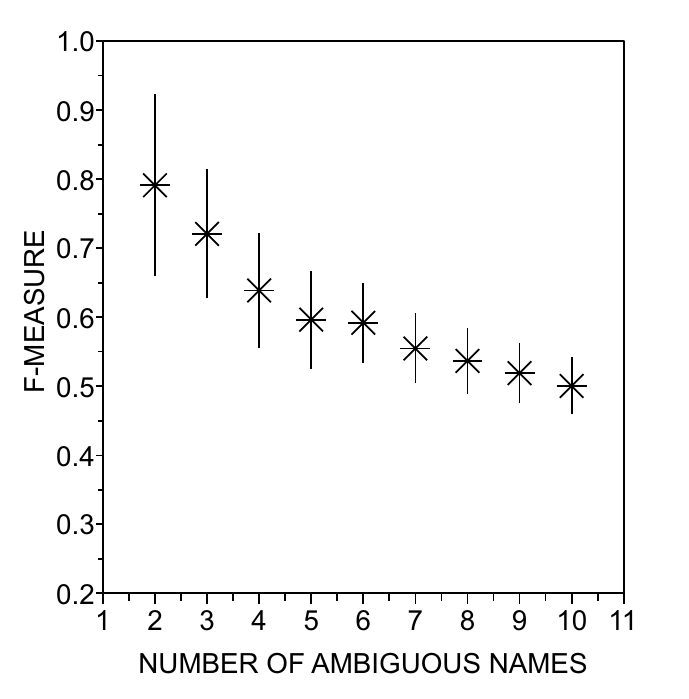}\label{fig:Regular-Lattice-lambda0}}}%
    \subfloat[]
    {{\includegraphics[angle=0, width=0.24\textwidth]{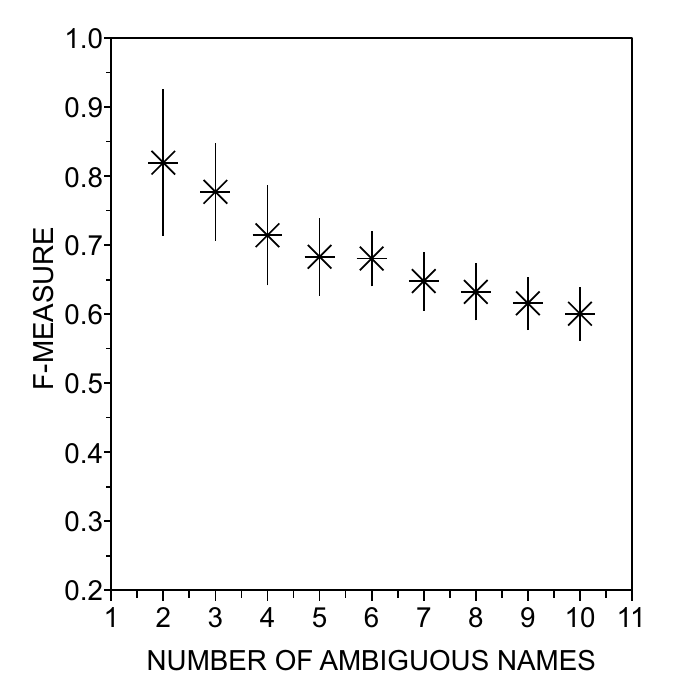}\label{fig:Regular-Lattice-lambda05}}}\\
    \caption{\label{fig.3} f-measure for the disambiguation task based on the topological approach. The algorithms used were (a) C4.5 and (b) $k$NN-1. In all cases, the ability to discriminate authors decreased as the number of ambiguous entities increased (Figure S2 of the SI brings the analogous curves for the other algorithms).}
\end{figure}

To our knowledge this is the first attempt to use the topology of collaborative networks for disambiguating authors. We used a set of $7$ topological measurements described in the methodology, but in principle other local measurements could have also been employed. The results in Figure \ref{fig.3} show that the overall discrimination ability using the network topology is worse than that obtained with the analysis of connectivity (see Figure \ref{fig.2}). However, the discrimination based on topological features was found to be statistically significant as depicted in {Table S2 of the SI, which points to authors exhibiting particular patterns of connectivity in collaborative networks.}

We also investigated which topological features were most efficient for discriminating authors, ranking them according to two criteria. The first criterion is based on the information gain~
achieved for each measurement and the second one is based on a methodology analogous to the Mann-Whitney U test~\cite{mann} (see SI for details regarding both methods). While the former has the advantage that the ranking is algorithm independent, it might overlook interactions between features because they are evaluated individually. For this reason, we also devised a methodology that does not ignore interactions between features. More specifically, the Mann-Whitney U test sorts the classifiers according to their accuracy rate. Then, the relevance of attributes is assigned according to their frequency in the top classifiers.\footnote{Even though the  Mann-Whitney U test relies on the pattern recognition strategy to perform the ranking, in our experiments all algorithms displayed practically the same results.}

The most efficient measurements for discrimination were $\langle k_n \rangle$ and $\langle s_n \rangle$, as shown in the ranking in Table~\ref{tab.rankey}, bringing the percentage of cases in which these measurements appeared as the best feature in the most efficient algorithm ($k$NN). The table also shows the corresponding $p$-value, considering a random ranking of measurements as null model, for both ranking criteria. Significantly, the clustering coefficient $C$ did not appear among the most important features for distinction. These results may be interpreted as follows. Discrimination appears to be governed by the average number of co-authors of neighbors, and to a lesser extent by the strength of such connections. Therefore, the most relevant information is actually the number of connections with external authors, i.e., the number of connections with authors who have never co-authored a paper with the author represented by the node under analysis. This means that the structure of the neighbors allows a better characterization than the local structure of the node itself, consistent with the findings from the analysis of connectivity.

\begin{table*}
\centering
\caption{\label{tab.rankey} \% of cases in which $\langle k_{n} \rangle$ and $\langle s_{n} \rangle$ appeared in the first position of the ranking performed using the Information Gain Criterion and the Mann Whitney test~\cite{mann} with the $k$NN algorithm. The $p$-value corresponds to the likelihood of the corresponding percentage to be obtained considering as null model a random ranking of measurements. \#$N$ is the number of ambiguous names for an author.}
\begin{tabular}{@{}|c|cc|cc|cc|cc|}
\hline
       & \multicolumn{4}{c|}{Information Gain Criterion} & \multicolumn{4}{c|}{Mann Whitney criterion} \\
\hline
 \#$N$    & \multicolumn{2}{c|} {\bf $\langle k_{n} \rangle$} & \multicolumn{2}{c|} {\bf $\langle s_{n} \rangle$}  & \multicolumn{2}{c|} {\bf $\langle k_{n} \rangle$} & \multicolumn{2}{c|} {\bf $\langle s_{n} \rangle$} \\
{\bf }	&	{\bf \%}	&	{\bf $p$-value}	&	{\bf \%}	&	 {\bf $p$-value}	& {\bf \%}	&	{\bf $p$-value}	&	 {\bf \%}	 &	 {\bf $p$-value}\\
\hline
Two		&	$56.8~\%$	&	$< 1.0 \times 10^{-15}$ &		$31.1~\%$	&	 $9.6 \times 10^{-5}$	 &		 $45.2~\%$	&	 $3.8 \times 10^{-14}$		 &	 $50.0~\%$	&	$< 1.0~10^{-15}$	\\		
Three	&	$67.4~\%$	&	$< 1.0 \times 10^{-15}$ &		$23.1~\%$	&	 $3.9 \times 10^{-2}$	 &		 $70.0~\%$	&	 $< 1.0 \times 10^{-15}$	&	 $29.5~\%$	&	$1.2~10^{-3}$	 \\
Four	&	$73.2~\%$	&	$< 1.0 \times 10^{-15}$ &		$20.6~\%$	&	 $7.0 \times 10^{-2}$	 &		 $73.7~\%$	&	 $< 1.0 \times 10^{-15}$	&	 $26.3~\%$	&	$2.1~10^{-2}$	 \\
Five	&	$73.2~\%$	&	$< 1.0 \times 10^{-15}$ &		$23.1~\%$	&	 $3.9 \times 10^{-2}$	 &		 $81.6~\%$	&	 $< 1.0 \times 10^{-15}$	&	 $18.4~\%$	&	$7.3~10^{-1}$	 \\
Six		&	$75.8~\%$	&	$< 1.0 \times 10^{-15}$ &		$20.6~\%$	&	 $7.0 \times 10^{-2}$	 &		 $91.1~\%$	&	 $< 1.0 \times 10^{-15}$	&	 $8.9~\%$	&	$\simeq 1.0$	 \\
Seven	&	$71.1~\%$	&	$< 1.0 \times 10^{-15}$ &		$25.8~\%$	&	 $1.0 \times 10^{-2}$	 &		 $86.3~\%$	&	 $< 1.0 \times 10^{-15}$ 	 &	 $13.2~\%$	&	$\simeq 1.0$	 \\
Eight	&	$66.3~\%$	&	$< 1.0 \times 10^{-15}$ &		$30.0~\%$	&	 $3.0 \times 10^{-4}$	 &		 $96.8~\%$	&	 $< 1.0 \times 10^{-15}$	&	 $3.2~\%$	&	$\simeq 1.0$	 \\
Nine	&	$69.4~\%$	&	$< 1.0 \times 10^{-15}$ &		$26.3~\%$	&	 $7.2 \times 10^{-3}$	 &		 $82.6~\%$	&	 $< 1.0 \times 10^{-15}$	&	 $17.4~\%$	&	$8.4~10^{-1}$	 \\
Ten		&	$42.6~\%$	&	$7.7 \times 10^{-13}$	 &		$52.6~\%$	&	 $< 1.0 \times 10^{-15}$&		 $92.1~\%$	&	 $< 1.0 \times 10^{-15}$	&	 $7.9~\%$	&	$\simeq 1.0$	 \\
\hline
\end{tabular}
\end{table*}

    \subsection{Connectivity and topology combined in a real network}


The strategies based on the connectivity (with 3rd hierarchy) and topological features were combined in the disambiguation task for a real database derived from the IMDb database for actors in movies (details are given in the SI). Table \ref{imdb_table} shows the accuracy rate achieved for each actor and the corresponding $p$-value (assuming as null model a random disambiguation system). For each actor, the scores shown were obtained with topological (TP) features, connectivity (CN) and with the combination of both strategies (CN + TP). For Attila and Matt Hughes, TP-features alone performed worse than CN-features, but the best result was reached with the strategies combined. Likewise, for Igor and Justin Long, the combination generated the best disambiguation. For Bill Balley, surprisingly, TP-features alone yielded the best results.
For Steve Austin, the combination also yielded the best result, although the same quality had already been obtained with CN-features. Finally, for Christian, even the random disambiguation system was already very efficient, and therefore a comparison is void. The results in this study are analogous to the findings in the task of recognizing authorship in written texts~\cite{interm}, in which the topology was proven useful for revealing patterns related to writing style.

\begin{table}
\centering
\caption{\label{imdb_table} Accuracy rate and $p$-value obtained for the disambiguation based on topological (TP) and connectivity (CN) measurements. The combination (CN+TP) of features was also examined. For all actors, the topological features appear in the best classifiers.}
\begin{tabular}{@{}|c|c|c|c|c|}
\hline
Actor & Classif.  & Acc.   &   $p$-value  &  Features \\
\hline
Attila       & C4.5        &   64.3 \%       &    $1.0 \times 10^{-1}$ & TP \\
             & kNN         &   78.6 \%       &    $2.8 \times 10^{-2}$ & CN \\
             & kNN         &   85.7 \%       &    $6.0 \times 10^{-3}$ & CN + TP \\
\hline
 Matt        & kNN         &   77.4 \%       &    $5.8 \times 10^{-1}$ & TP \\
 Hughes      & kNN         &   80.7 \%       &    $8.5 \times 10^{-2}$ & CN \\
             & kNN         &   83.9 \%       &    $3.6 \times 10^{-2}$ & CN + TP \\
\hline
             & RIP.      &   81.8 \%       &    $5.5 \times 10^{-2}$ & TP    \\
Igor         & kNN         &   81.8 \%       &    $5.5 \times 10^{-2}$ & CN    \\
             & kNN         &   86.4 \%       &    $1.8 \times 10^{-2}$ & CN+TP \\
\hline
Justin       & kNN         &   95.8 \%       &    $2.2 \times 10^{-4}$ & TP    \\
 Long        & kNN         &   93.6 \%       &    $3.0 \times 10^{-2}$ & CN    \\
             & RIP.      &   95.8 \%       &    $2.2 \times 10^{-4}$ & CN+TP \\
\hline
Bill         & C4.5 & 96.0 \% &  $2.2 \times 10^{-2}$ & TP\\
     Bailey  & kNN  & 87.8 \% &  $6.1 \times 10^{-1}$ & CN \\
             & kNN  & 87.8 \% &  $6.1 \times 10^{-1}$ & CN+TP\\
\hline
Steve        & Bayes &   84.8 \%       &    $1.0 \times 10^{-1}$ & TP \\
 Austin      & Bayes &   88.9 \%       &    $2.2 \times 10^{-2}$ & CN \\
             & kNN         &   88.9 \%       &    $2.2 \times 10^{-2}$ & CN + TP\\
\hline
             & kNN         &   97.1 \%       &    $4.5 \times 10^{-1}$ & TP \\
Christian    & kNN         &   97.6 \%       &    $2.9 \times 10^{-1}$ & CN \\
             & kNN         &   97.6 \%       &    $2.9 \times 10^{-1}$ & CN + TP\\
\hline
\end{tabular}
\end{table}



\section{Conclusion} \label{conclusao}

	Two innovative approaches were introduced in this paper for disambiguating names in collaborative networks. In the first, we extended the traditional method based on the connectivity with immediate neighbors in networks by incorporating the analysis of higher hierarchies. We showed that the 3rd hierarchy leads to a considerably improved performance in the disambiguation task. In the second approach, we used for the first time – to our knowledge – the topology of networks for disambiguating names. The two most efficient measurements for distinguishing authors were $\langle k_n \rangle$ and $\langle s_n \rangle$, i.e. distinction depends mainly on the connectivity of the neighbors. This reinforces the importance of considering deeper hierarchies while analyzing collaborative networks~\cite{amancio}.

All of these results were obtained for a subnetwork from the arXiv repository for the area of complex networks, in which ambiguity was deliberately introduced. The option for this artificial system was made to ensure a reliable dataset and the statistical significance of our analysis. Furthermore, the robustness of the analysis was ensured by employing various pattern recognition methods, belonging to both supervised and unsupervised machine learning paradigms, with which similar results were obtained. Obviously, the innovative approaches can be extended to real networks, and indeed we showed that for a network of movie actors. In particular, we noted that combination of the two approaches leads to improved performance in disambiguating names, which is promising for further applications requiring removal of ambiguity in databases.



For future works, we intend to further investigate if the hierarchical characterization introduced in the traditional analysis can further improve the ability of discrimination in the topological characterization. Also, we intend to analyze the performance of similarity measures based on complex networks, such as the Katz similarity~\cite{net8}. {Another point of future investigation concerns the verification of the precise influence of sampling on the topological analysis, because incomplete databases usually generate worst disambiguating systems (see SI).}
Finally, we plan to apply the topological approach to the problem of disambiguating words in written texts (word sense disambiguation)~\cite{epldiego}.


\acknowledgments
The authors would like to acknowledge CNPq (Brazil) and FAPESP (2010/00927-9) for the sponsorship.

\end{document}